\documentclass[aps,pre,reprint,amsmath,amssymb,floats,longbibliography,superscriptaddress]{revtex4-2}

\usepackage{graphicx}
\usepackage{bm}
\usepackage{newunicodechar}
\usepackage[utf8]{inputenc}
\usepackage[T1]{fontenc}
\newunicodechar{ȩ}{\k{e}}
\usepackage{color}

\begin{document}
\title{Suppression of discontinuous phase transitions by particle diffusion}
\author{Chul-Ung Woo}
\affiliation{Department of Physics, University of Seoul, Seoul 02504, Korea}
\author{Heiko Rieger}
\affiliation{Department of Theoretical Physics \& Center for Biophysics,
Saarland University, 66123 Saarbr\"ucken, Germany}
\affiliation{INM – Leibniz Institute for New Materials, Campus D2 2, 66123
Saarbrücken, Germany}
\author{Jae Dong Noh}
\affiliation{Department of Physics, University of Seoul, Seoul 02504, Korea}
\date{\today}

\begin{abstract}
    We investigate the phase transitions of the $q$-state Brownian Potts 
    model in two dimensions (2d) comprising Potts spins that diffuse
    like Brownian particles and interact ferromagnetically with other spins 
    within a fixed distance. With extensive Monte Carlo simulations
    we find a continuous phase transition from
    a paramagnetic to a ferromagnetic phase even for $q>4$. This is in 
    sharp contrast to the existence of a discontinuous phase transition
    in the equilibrium $q$-state Potts model in 2d with $q>4$. We present
    detailed numerical evidence for a continuous phase transition and
    argue that diffusion generated dynamical positional disorder
    suppresses phase coexistence leading to a continuous transition.
\end{abstract}

\maketitle

\section{Introduction}\label{sec:intro}
Phase transitions and critical phenomena have long been studied in
statistical physics. Owing to advances in theoretical and numerical
methods, equilibrium phase transitions are quite well understood. 
Microscopically different many-body models can have the same 
critical exponents characterizing the variation of certain physical 
properties near a critical point, thus representing the same 
universality class, which depend on but a few determinants 
like order parameter symmetry, spatial dimensionality, presence
of quenched disorder etc.~\cite{Stanley.1999}. Recently, nonequilibrium 
phase transitions attracted growing
interest~\cite{Hinrichsen.2000,Wittkowski.2014}. 
Broken detailed balance distinguishes nonequilibrium from equilibrium
systems. As the detailed balance can be broken in many ways, 
nonequilibrium systems display a larger variety of phase transition 
scenarios.
%, the universal features of which are less well understood yet. 

A now paradigmatic example is active matter, which is characterized 
by energy consuming, self-propelled constituents and therefore driven 
out of equilibrium. Self-generated motility is responsible for various 
collective phenomena, like flocking, motility induced phase separation, 
active turbulence etc.~\cite{Ramaswamy.2017,Shaebani.2020,OByrne.2022}.
The seminal paper by Vicsek {\em et al.}~\cite{Vicsek.1995} demonstrated
that motility stabilizes long range orientational order in 2d,
which would be unstable in the corresponding 2d equilibrium systems 
according to the Mermin-Wagner theorem~\cite{Goldenfeld.1992}. 
Motility can also induce phase separation of active particles with
repulsive interactions~\cite{Cates.2014}. 
There are surging research
activities to unveil the role of the motility in many-body
systems~\cite{Solon.2013,Solon.2015f1h,Chatterjee.2020,Mangeat.2020,Shaebani.2020}.

Many studies of ensembles of self-propelled particles focus
on the collective behavior of the spatial degrees of freedom such 
as the velocity and the position of particles.
Motivated by collective effects that are induced by motility
we focus in this paper on spins which are not fixed but move in space
and search for an order-disorder phase transition. 
Concretely, we study the $q$-state Brownian Potts
model in 2d, represented by Potts spins that diffuse like Brownian 
particles and interact ferromagnetically with other spins 
within a fixed distance.
Whereas in the active Ising \cite{Solon.2013,Solon.2015f1h}
or Potts \cite{Chatterjee.2020,Mangeat.2020,Shaebani.2020} model
the spin of a particle determines its direction of motion such that spin and
spatial degrees of freedom are mutually coupled, 
we focus here on a uni-directional coupling: particles diffuse 
{ \it freely irrespective of spin states} 
while a spin interaction network evolves in time as 
particles diffuse. Even this simplified model displays interesting 
critical phenomena due to the ferromagnetic interactions.

The paper is organized as follows:
In Sec.~\ref{sec:model}, we introduce the $q$-state Brownian Potts model and
discuss its features that are different from the equilibrium Potts model in 2d.
In Sec. III we present our results of extensive Monte Carlo simulations, which show that the $q$-state Brownian Potts model has a continuous phase transition for all values of $q$, and determine the critical exponents. In Sec. IV we present an argument, based on a comparison of the time scales for particle diffusion and spin-spin correlation propagation, that diffusion impedes phase coexistence, which 
renders the transition continuous. We conclude the paper with summary and
discussions in Sec.~\ref{sec:summary}.

\section{$q$-state Brownian Potts model}\label{sec:model}
\begin{figure}[ht]
  \includegraphics*[width=0.8\columnwidth]{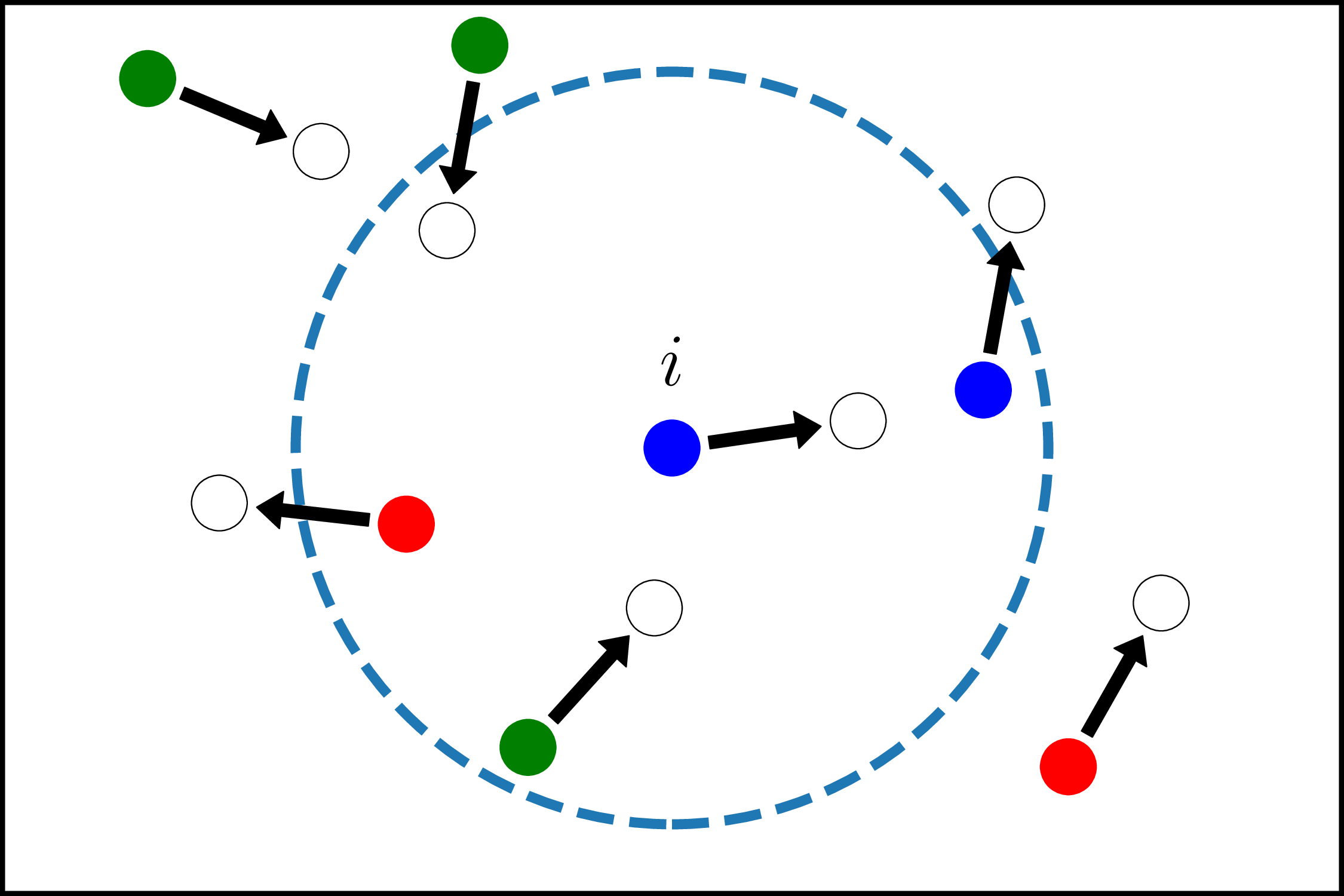}
    \caption{{Schematic diagram illustrating the dynamic
            rule. The filled symbols represent the Brownian particles whose
            Potts spin states are color-coded.
            Each particle interacts ferromagnetically with the others within
            the distance $r_0$.  %flips its Potts spin state as a result of the
            %ferromagnetic spin interaction of range $r_0$.
            For instance, the dashed circle denotes the interaction range of
            the particle $i$.
            After spin flips, particles perform 
            a random jump of length $v_0$ to a next position denoted by
    open symbols.}}
    \label{fig1}
\end{figure}

The model consists of $N=\rho L^d$ particles of density $\rho$ 
on a square of area $L^2$ with periodic boundary conditions. 
The position of particle $i$ is denoted as 
$\bm{r}_i \in \mathbb{R}^2$ and its Potts spin state by $\sigma_i \in
\{1,\cdots, q\}$. The spins interact ferromagnetically with
other particles $j$ in a distance 
$\vert {\bf r}_j-{\bf r}_i\vert\le r_0\equiv 1$.
We adopt parallel update dynamics in discrete time units: 
Given $\{\sigma_i(t), \bm{r}_i(t)\}$ at time step $t$,
the spin states of all particles are updated according to the probability
\begin{equation}
P(\sigma_i(t+1)=\sigma) = \frac{1}{Z} \exp\left[K\sum_{j\in \mathcal{N}_i}
\delta(\sigma,\sigma_j(t))\right],
\label{spin_flip}
\end{equation}
where $\delta(a,b)$ is the Kronecker-$\delta$ symbol, $\mathcal{N}_i$ denotes
the set of particles within the distance $r_0$ from particle $i$, $K>0$
the ferromagnetic spin-spin interaction strength, and $Z$ is a
normalization constant. 
Subsequently each particle performs a jump of length $v_0$ in
a random direction, $\bm{r}_i(t+1) = \bm{r}_i(t) + v_0
(\cos\theta_i,\sin\theta_i)$, where $\theta_i\in(-\pi,\pi]$ 
are uniformly distributed independent random variables,
generated anew in each time step~({see Fig.~\ref{fig1}}).
Note that the heat bath algorithm for the ferromagnetic $q$-state Potts
model~\cite{Wu.1982k4e}
has the same spin flip probability as in Eq.~\eqref{spin_flip}. Thus, when
$v_0=0$, our model becomes the equilibrium $q$-state Potts
model on a random lattice~\footnote{Since we adopt 
    the parallel update rule, the spins relax
into a thermal equilibrium state associated with a Hamiltonian that is slightly
modified from the conventional Potts model Hamiltonian.}.

Particle diffusion introduces interesting features. It breaks the detailed
balance and drives the system out of equilibrium. 
Since particles diffuse freely the system can be seen as being 
in contact with two thermal heat baths: 
a finite temperature heat bath for the spin degrees of freedom with
temperature $T \propto 1/K$
and an infinite temperature heat bath for the spatial degree of freedom. 
Nonequilibrium phase transitions of
systems in thermal contact with two heat baths were investigated, for instance, 
in the context of the random $q$-neighbor Ising
model~\cite{Park.2017,Jedrzejewski.2015}. 

Particle diffusion leads to particle density fluctuations, which 
acts like time-dependent disorder in the spin-spin interactions. 
Due to hopping, spin-spin interactions are spatially heterogeneous and
each spin interacts with a fluctuating number of neighboring spins. 
Particle hoppings and spin flips occur simultaneously. Thus, the microscopic 
time scales for the particle hopping $\tau_p$ and for the spin flip 
$\tau_s$ are comparable.  
One can predict the consequence of the time-dependent disorder 
if the two time scales are well separated. For instance, in the
limit $\tau_s/\tau_p\to 0$ or $v_0\to0$, the time-dependent disorder 
becomes quenched. There are rigorous results predicting 
the "rounding" of a discontinuous phase transition  by quenched 
disorder for low dimensional 
system~\cite{Imry.1975,Greenblatt.2010,Aizenman.1989,Hui.1989}.
Thus, in the limit $v_0\to 0$, we expect that our model belongs to the 
universality class of the Potts model with quenched disorder
and undergoes a continuous phase transition at all values of
$q$~\cite{Olson.1999}. In the opposite limit, $\tau_s/\tau_p\to\infty$ 
or $v_0\to\infty$, we expect that the mean field
theory is valid because spins are well mixed. 
The universality class of the transition in the intermediate case, 
$0<v_0<\infty$, remains elusive, which is what we intend to clarify in this paper. Since the two microscopic time scales are comparable, the macroscopic time scales 
at which particle density fluctuations and spin fluctuations propagate
should be compared, which we will do in Sec.~\ref{sec:nature}.

It is worth mentioning the active Ising and the active Potts 
model introduced in
Refs.~\cite{Solon.2013,Solon.2015f1h,Chatterjee.2020,Mangeat.2020}.
In these models, particles with spin move on a 2d lattice
and interact ferromagnetically with other particles on the same site. 
Their motion is biased towards a direction determined by their spin state. 
These models are discretized versions of the 
flocking model of Ref.~\cite{Vicsek.1995}. 
Due to the bias and the ferromagnetic interaction, particles tend to 
move persistently together with neighboring particles, which stabilizes 
collective motion denoted as flocking.
In contrast to the active Ising and Potts models, particles in the Brownian
Potts model diffuse freely irrespective of the spin states. 
Thus, our model may be regarded as a passive Potts model. 

\section{Numerical Results}\label{sec:numerical}
We have performed extensive Monte Carlo simulations for the Brownian Potts 
model with $q=2,\cdots, 8$. 
Starting from a random initial state, Monte Carlo simulations are 
performed up to $T_{max}$ time steps and a time series of the fraction
$n_\sigma$ of particles in the spin state $\sigma~(=1,\cdots, q)$ is
obtained. 
The Potts order parameter~\cite{Wu.1982k4e}, quantifying ferromagnetic 
order, is 
\begin{equation}
    m_s = \left( q \max_{\sigma} \{n_\sigma\} - 1\right)/(q-1) .
\end{equation}
Potts spin states can be represented by unit vectors 
$\{ \bm{e}_\sigma\}$ in the $(q-1)$-dimensional space~\cite{Wu.1982k4e}. 
Using this representation, one can also define a vector order parameter
\begin{equation}
    \bm{m}_v = \frac{1}{N} \sum_{i=1}^N \bm{e}_{\sigma_i} = \sum_\sigma
    n_\sigma \bm{e}_\sigma.
\end{equation}
Using the time series $\{n_\sigma\}$, 
we can calculate the steady state average of $m_s$ and $|\bm{m}_v|$, and
their moments and probability distributions.
Both quantities display a qualitatively similar behavior.
We investigate the phase transition by varying
the coupling strength $K$ with fixed
interaction range $r_0 = 1$, hopping length $v_0 = 1/2$, and 
particle density $\rho=1$.
{We also performed the simulation study at $\rho=2$ and $4$ 
    and obtained the same conclusion. Thus we only present the numerical
results at $\rho=1$.}
The largest system size we considered is $L = 512$ with $T_{max} = 4\times 10^8$, 
which is sufficiently long to reach a steady state. 
Data in the time interval from $T_{max}/10$ to $T_{max}$ steps 
are used for the steady state average.
In order to estimate the statistical uncertainty of the numerical results, 
we used the bootstrap or resampling method~\cite{newman1999monte}: Given a
steady-state time series $\{n_\sigma\}$, we resample $S$ 
subsets each of which consists of randomly chosen $T_{max}/S$
data points. A statistical error of a quantity is measured by the standard
deviation of $S$ sampled averages. We chose $S=100$.

\begin{figure}[ht]
  \includegraphics*[width=\columnwidth]{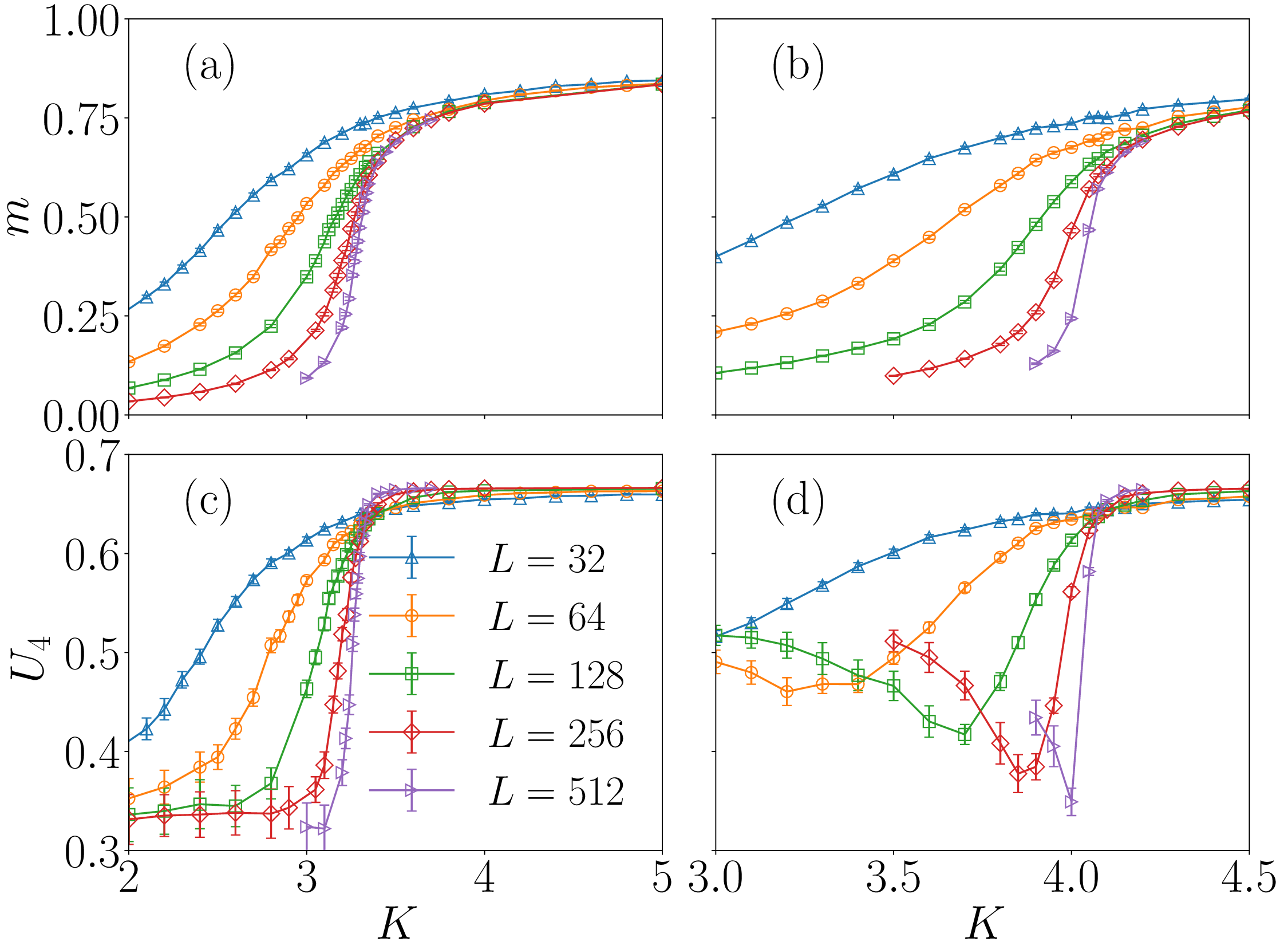}
    \caption{Order parameter (top) and the Binder cumulant (bottom)
    of the Brownian Potts model for $q=3$ (a,c) and $q=6$ (b,d).}
    \label{fig2}
\end{figure}

Figure~\ref{fig2} shows the order parameter $m = \langle
|\bm{m}_v|\rangle$ and the Binder cumulant
$U_4 \equiv 1-\langle |\bm{m}_v|^4\rangle / (3\langle |\bm{m}_v|^2\rangle^2)$ for
$q=3$ and $6$.
The order parameter shows that the system has 
a phase transition from a paramagnetic to a ferromagnetic phase.
When $K$ is smaller than a critical value $K_c$, 
the order parameter decays to zero as $L$ increases. 
On the other hand, when $K > K_c$, it converges to a finite value.
The critical interaction strength $K_c$ can be estimated from 
the intersection of the curves of the Binder cumulants 
$U_4$ for different system sizes (see Fig.~\ref{fig2}(c) and (d)). 
We obtain $K_c(q=3) = 3.33(3)$ and $K_c(q=6) = 4.08(3)$. 
We will discuss the critical exponents below.

\subsection{Order of the transition}
The Binder cumulant $U_4$ shown in Fig.~\ref{fig2}(c) and (d) has a dip
in the paramagnetic phase. The dip becomes more pronounced at larger 
values of $q$.
The Binder cumulant diverges (to $-\infty$) at a discontinuous phase 
transition point~\cite{Binder.1997}.
Recalling that the equilibrium Potts model in 2d
has a discontinuous phase transition for $q>4$, we need to 
examine whether the dip is an indication of the discontinuous phase transition.

\begin{figure}[ht]
  \includegraphics*[width=\columnwidth]{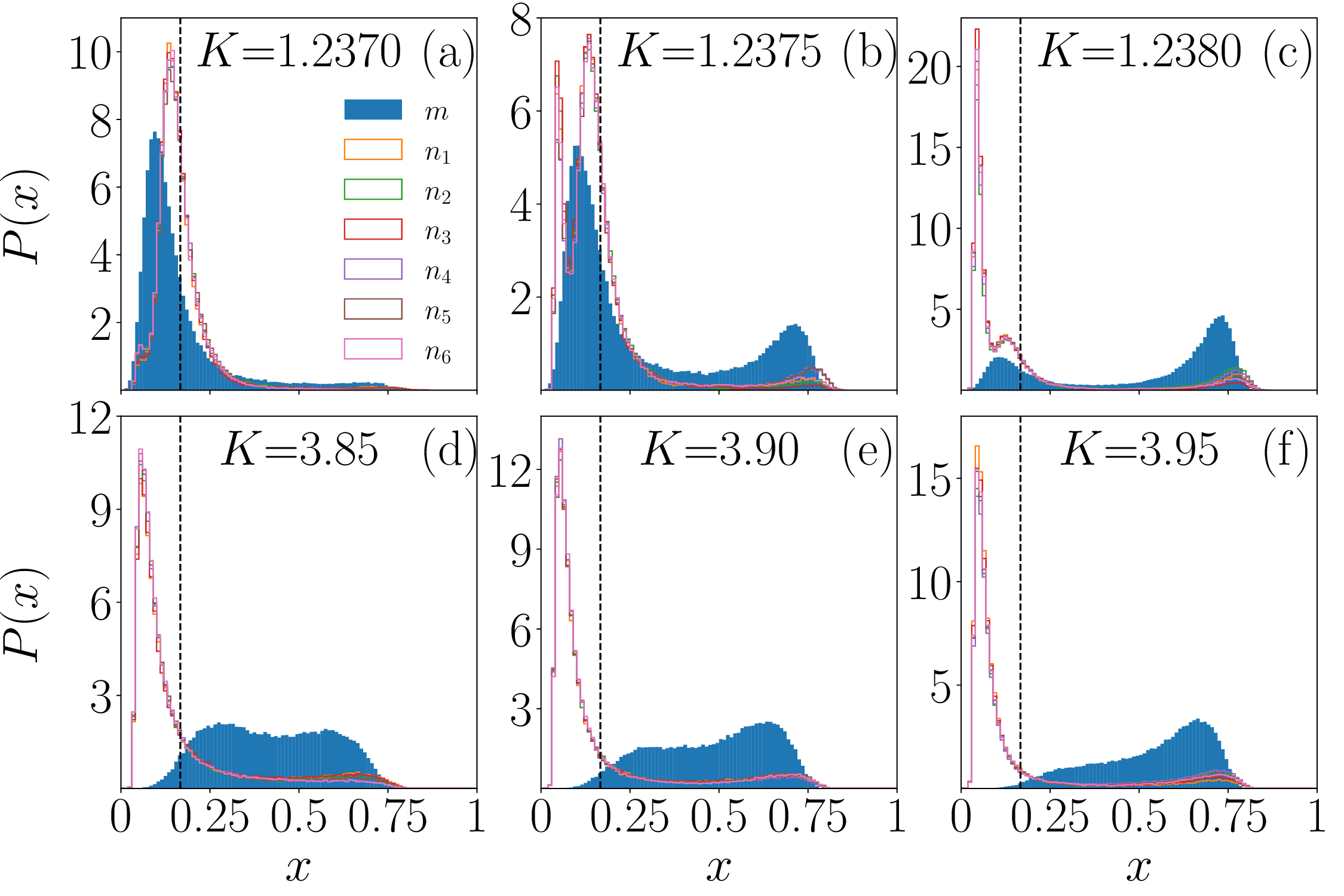}
  \caption{Histograms of the order parameter $x = |\bm{m}_v|$~(blue, filled curves) 
    and spin fractions $x = n_\sigma$ with $\sigma=1,2,\cdots, q$~(solid lines). 
    Top row (a,b,c) shows the histograms for the equilibrium Potts model,
    bottom row (d,e,f) for the Brownian Potts model. Data are for $q=6$, 
    $L=128$ and $K\simeq K_c$. 
    Note that $K_c = \ln(1+\sqrt{6}) \simeq 1.2382$ for the 
    equilibrium 6-states Potts model and $K_c \sim 4.08(3)$ for the 
    Brownian Potts model. The dotted vertical lines indicate the value $x=1/q$.}
  \label{fig3}
\end{figure}

%Phase coexistence at the discontinuous phase transition can be detected with
%the order parameter histogram.
In Fig.~\ref{fig3}, we compare the order parameter histograms of the equilibrium 
Potts model and the Brownian Potts model with $q=6$. 
The equilibrium Potts model displays a behavior characteristic for 
a discontinuous phase transition: The histogram has a single 
peak corresponding to a paramagnetic phase at small $K$~(Fig.~\ref{fig3}(a)), 
double peaks indicating phase coexistence~(Fig.~\ref{fig3}(b)) for intermediate
values of $K$, and a single peak corresponding to a
ferromagnetic phase at large $K$~(Fig.\ref{fig3}(c)). 
The order parameter histogram for the Brownian Potts model also has double
peaks, but they are much less pronounced than those observed in the 
equilibrium model. 

The order parameter histograms, $P(|\bm{m}_v|)$, alone do not provide a
conclusive evidence for phase coexistence. Thus, we further
investigate the histograms, $P(n_\sigma)$, of the fraction $n_\sigma$ 
of particles in the spin state $\sigma~(=1,\cdots, q)$. 
In the disordered paramagnetic phase, all spin states are equally populated
with statistical fluctuations. Namely, all histograms of $n_\sigma$ 
should have a peak around $x=1/q$. In the ordered 
ferromagnetic phase, one spin state, say $\sigma_m$, dominates. Thus, the
histogram of $n_{\sigma_m}$ should have a peak at $x>1/q$ while the other
$(q-1)$ histograms at $x < 1/q$. 
In Fig.~\ref{fig3}(b) for the equilibrium Potts model, 
we find the peaks corresponding to both phases.
The central peak at $x\simeq 1/q$ corresponds
to the disordered phase, which is well separated from the other two peaks
corresponding to the ordered phase. Our simulation time $T_{max}$ is so 
long that the system alternates between the disordered state and the ordered
states with different $\sigma_m$. This three-peak structure is an 
evidence for phase coexistence in the equilibrium 6-state Potts model.

On the other hand, the histograms of $n_{\sigma}$ for the Brownian Potts model
do not have a peak at $x \simeq 1/q$ representing the paramagnetic 
phase~(see Fig.~\ref{fig3}(d,e,f)). 
This is a clear and decisive evidence for the absence of phase coexistence. 
We also performed the same analysis for other values of $q=2,\cdots, 8$,
which lead to the same conclusion that the Brownian Potts model has a
continuous phase transition irrespective of the value $q$. 
We will substantiate this conclusion with a theoretical argument later.

The apparent double peaks in the order parameter histogram is attributed to
the discrete symmetry of the Potts spin. The Potts order parameter $\bm{m}_v$
lies within a $(q-1)$-dimensional polyhedron~\cite{Wu.1982k4e}. For
instance, it lies inside an equilateral triangle for $q=3$.
As the coupling constant approaches $K_c$ from below, the order parameter
deviates from the center of the polyhedron and moves toward a
vertex developing a peak in the order parameter histogram. 
Due to fluctuations at finite $L$, it does not stay 
near a single vertex but keeps diffusing to the other vertices. 
In the transient period, the order parameter magnitude shrinks because it
is limited by the faces of the polyhedron. This
effect can result in an additional bump in the order parameter 
histogram. The broad double peaks observed in Figs.~\ref{fig3} (d) and (e) 
{ and the apparent dips in the Binder cumulant shown in Fig.~\ref{fig2} (d)}
are plausibly a consequence of this effect.

We also confirmed numerically that the equilibrium $(q=3)$-state Potts model 
in 2d has a similar bump in the order parameter histogram and a dip in the
Binder parameter. The equilibrium $3$-state Potts model 
is known to have a continuous phase transition.
Therefore, the unusual behavior of the order parameter histogram and the
Binder parameter should not be taken as an evidence for a
discontinuous phase transition. 
%They reflect the discrete order parameter symmetry. 

\subsection{Critical scaling of the order parameter}
\begin{figure}
    \includegraphics*[width=\columnwidth]{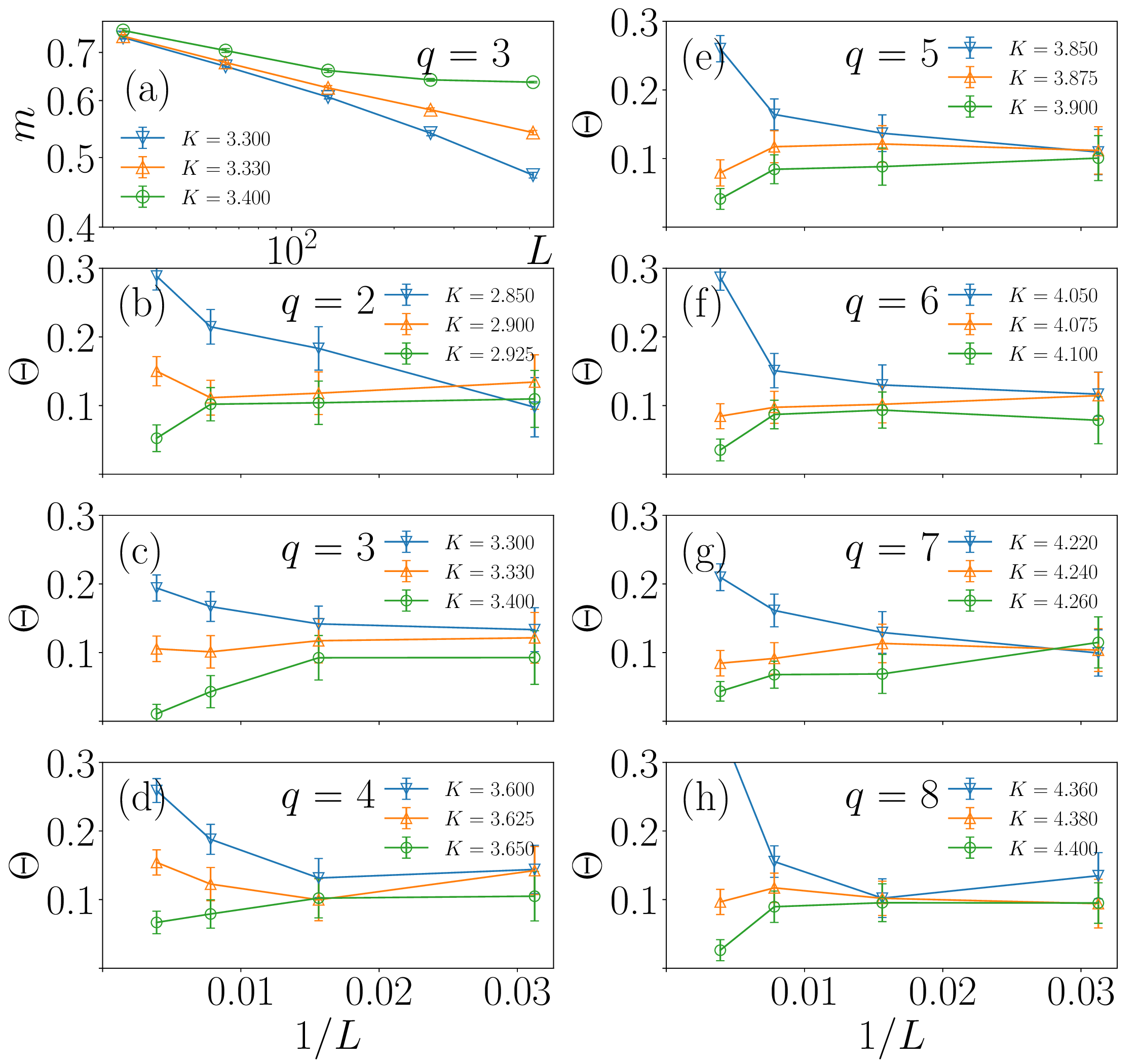}
    \caption{(a) Order parameter of the Brownian Potts model 
    	as function of $L$ for $q=3$ at different 
    	coupling constants $K$.
    (b-h) Effective exponent $\Theta$ as a function of $1/L$ 
    near the critical points for different values of $q$.}
    \label{fig4}
\end{figure}

After establishing that the phase transitions are continuous ones, 
we will estimate the critical exponents via finite size scaling. 
Near the critical point, the order
parameter is assumed to have the scaling form
\begin{equation}
    m(K,L) = L^{-\beta/\nu} F_m(t L^{1/\nu}),
    \label{m_scaling}
\end{equation}
where $t \equiv (K-K_c)/K_c$ is the reduced coupling strength  
(akin to a reduced temperature), $\nu$
is the correlation length exponent describing the divergence of the
correlation length $\xi \sim |t|^{-\nu}$, $\beta$ is the order parameter
exponent, and $F_m(x)$ is a scaling function. At the critical point, the
order parameter follows the power law $m(K_c,L) \sim L^{-\beta/\nu}$. 
The exponent $\beta/\nu$ can be estimated from the effective exponent 
$\Theta(K,L) \equiv -\log[m(K,2L)/m(K,L)]/\log{2}$. It converges to
$\beta/\nu$ at $K=K_c$ and crosses over to the trivial value $0$ for $K>K_c$
and $d/2$ for $K<K_c$. 
Figure~\ref{fig4}(a) demonstrates the power law scaling at the critical
point and the crossover in the off-critical regime.
In Figs.~\ref{fig4}(b-h), we present the effective exponent $\Theta$
as a function of $1/L$ near critical points at $q=2,\cdots, 8$. 
An overall curvature in the plot $\Theta$ vs. $1/L$ indicates a deviation from 
the critical point, which allows us to estimate the critical point 
and the exponent $\beta/\nu$
and their numerical uncertainty. They are summarized in Table~\ref{table1}.
The exponent $\beta/\nu$ falls within the range
$\beta/\nu = 0.10 \pm 0.05$ for all values of $2\leq q\leq 8$. 
From our data with $L\leq 512$, 
it is hard to draw a conclusion whether the exponent has a
$q$-dependence or not.

\begin{figure}
    \includegraphics*[width=\columnwidth]{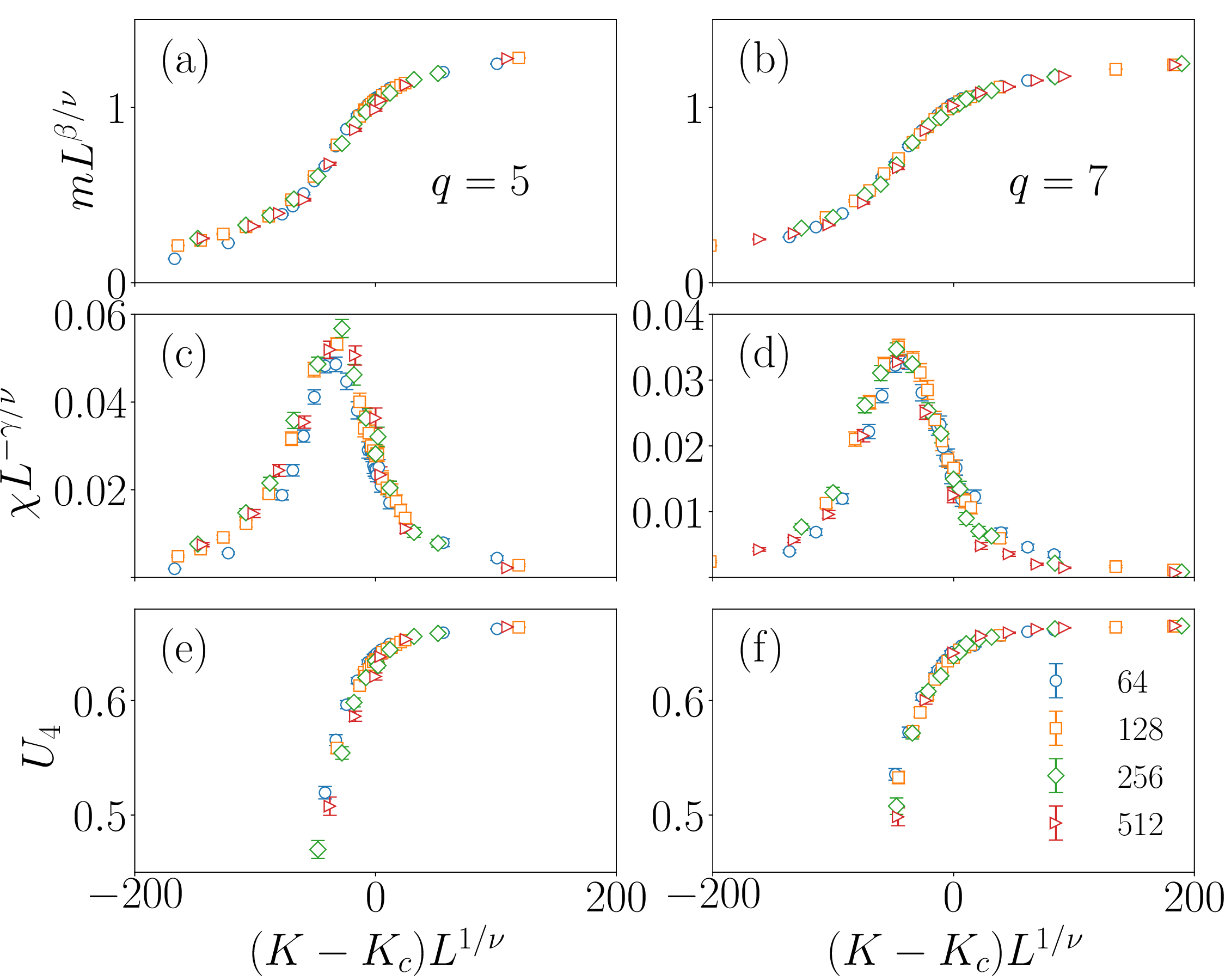}
    \caption{{Scaling plots of $m, \chi$, and $U_4$ at $q=5$ in (a), (c), and (e) and at
    $q=7$ in (b), (d) and (f) using the numerical estimates listed in
    Table~\ref{table1}.}}
    \label{fig5}
\end{figure}

In addition to the order parameter, we also calculated the
order parameter fluctuation $\chi \equiv N \left( \langle |\bm{m}_v|^2
\rangle - \langle |\bm{m}_v|\rangle^2\right)$. Analogously to the
magnetic susceptibility in the equilibrium system, it is expected to follow
the finite size scaling form
\begin{equation}
    \chi(K, L) = L^{\gamma/\nu} F_{\chi}(tL^{1/\nu}),
    \label{chi_scaling}
\end{equation}
where the exponent $\gamma$ characterize the power law scaling of 
$\chi \sim |t|^{-\gamma}$ and $F_\chi$ is a scaling function.
We have performed the data collapse analyses of $m$ and $\chi$ using the
scaling forms in Eqs.~\eqref{m_scaling} and \eqref{chi_scaling} to determine
the critical exponents $1/\nu$ and $\gamma/\nu$. 
Our estimates of the critical coupling constants and the critical 
exponents at all values of $q$ are summarized in Table~\ref{table1}.
Figure~\ref{fig5} shows the scaling plots at $q= 5$ and $7$. One observes that
$\chi$ suffers from strong finite size effects. 

Particles in the active Ising model~\cite{Solon.2015f1h} become passive
if one turns off the spin-dependent hopping bias. The phase transition in
that limit belongs to the universality class of the 2d
equilibrium Ising model~\cite{Solon.2015f1h}. Our result for $\beta/\nu$
is comparable with $(\beta/\nu) = 1/8, 2/15$ and $1/8$ of the lattice
$q$-state Potts model with $q=2, 3,$ and $4$, respectively, within the
numerical uncertainty. It may suggest that the Brownian Potts model belong
to the same universality class of the equilibrium Potts model.
However, the continuous phase transition in the
Brownian Potts model for $q>4$ excludes such a possibility.
{
The numerical results for $1/\nu$ at $q=3$ and $4$ are not consistent with 
those of the equilibrium 2d Potts model~[$1/\nu= 6/5~(q=3)$ and 
$3/2~(q=4)$] either~\cite{Wu.1982k4e}.
}

\begin{table}
    \caption{Critical interaction strength and the exponents.}
    \label{table1}
    \begin{ruledtabular}
        \begin{tabular}{c c c c c c}
%\hline \hline
 $q$ & $K_{c}$ &$\beta/\nu$ & $1/\nu$ & $\gamma/\nu$ &$Z_R$\\ [0.5ex]
 \hline
 2 & $2.90(5)$ & $0.10(5)$ & $1.00(10)$ & $1.70(20)$ & 2.15(10)\\
 3 & $3.33(3)$ & $0.10(5)$ & $1.03(10)$ & $1.75(20)$ & 2.25(10)\\
 4 & $3.63(3)$ & $0.10(5)$ & $1.05(10)$ & $1.80(20)$ & 2.35(10)\\
 5 & $3.87(5)$ & $0.10(5)$ & $1.08(10)$ & $1.85(20)$ & 2.35(10)\\
 6 & $4.08(3)$ & $0.10(5)$ & $1.10(10)$ & $1.90(20)$ & 2.38(10)\\
 7 & $4.24(3)$ & $0.09(5)$ & $1.13(10)$ & $1.95(20)$ & 2.40(10)\\
 8 & $4.38(3)$ & $0.09(5)$ & $1.15(10)$ & $1.95(30)$ & 2.45(10)\\
%\hline \hline
\end{tabular}
\end{ruledtabular}
\end{table}

\section{Time scales and nature of the transition}\label{sec:nature}
Density fluctuations due to diffusion render a particle-particle interaction 
network inhomogeneous. Particles in a dense~(dilute) region have more~(less)
neighbors. Moreover, the interaction network evolves in time as particles
diffuse. In order to gain an insight on the influence of the time-dependent 
disorder on the phase transition, we compare time scales for relevant degrees
of freedom. 

\begin{figure}[t]
  \includegraphics*[width=\columnwidth]{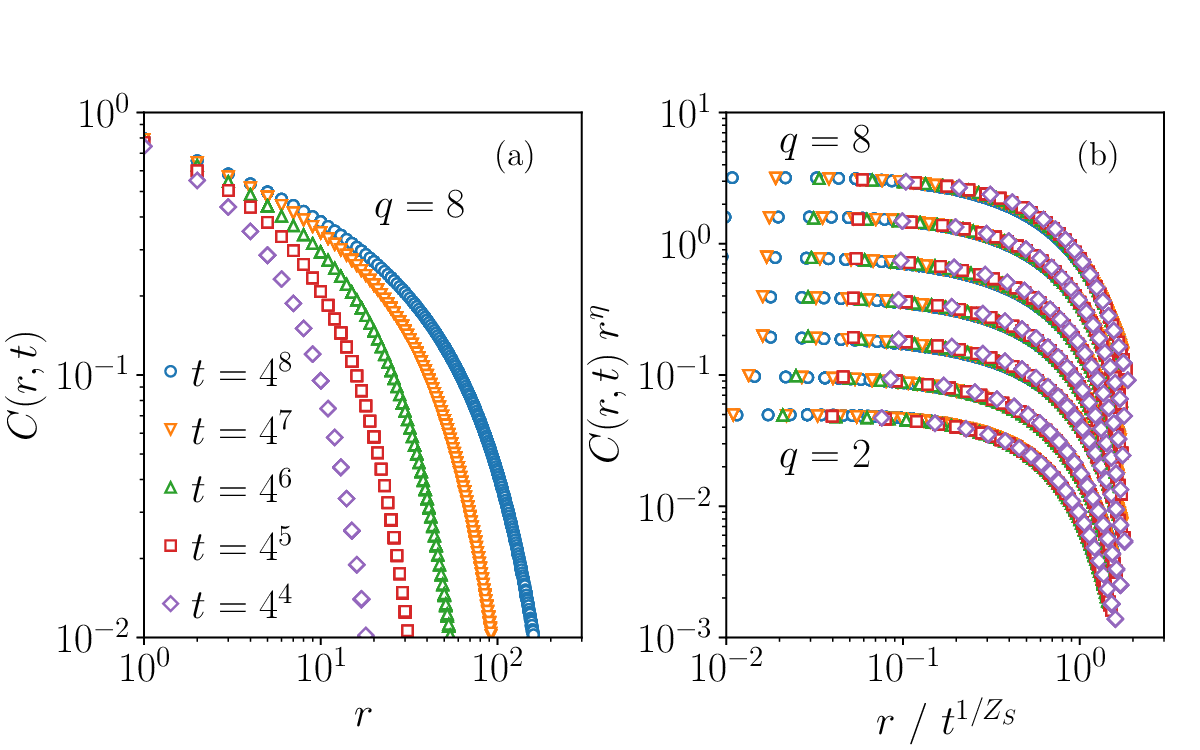}
  \caption{(a) Spin-spin correlation function at $t=4^4,\ldots, 4^8$
    evaluated at the critical point for $q=8$.
    Data points are aligned along a single curve,
    which indicates that the correlation function is isotropic.  (b) Scaling
  plots according to Eq.~\eqref{C_scaling} at all values of $2\leq q \leq 8$.
For a better visualization, we shift the data sets vertically by multiplying
$2^{q-6}$. All the data are obtained from the ensemble average over 100
samples. The system size is $L=4096$.} 
    \label{fig6}
\end{figure}

Obviously, there is a {\em diffusion time} scale $\tau_D \sim
\xi^{Z_D}$ with $Z_D = 2$ characterizing the particle diffusion over the
distance $\xi$. 
In addition, there is a {\em relaxation time scale} $\tau_R$
which it takes for the spins to reach the steady state.
We characterize the scaling behavior of $\tau_R$ 
from the {\em equal-time} spin-spin correlation function. 
For the correlation function of the off-lattice system, we divide the
two-dimensional plane into $L^2$ unit cells $\{\alpha = 1,\cdots,L^2\}$ 
and define a cell spin $\bm{S}(\bm{r}_\alpha, t) = \sum_{ \bm{r}_i \in \alpha}
\bm{e}_{\sigma_i}$ as the sum of spins of particles in a cell located at 
$\bm{r}_\alpha\in \mathbb{Z}^2$ at time $t$.
The equal-time correlation function is then defined as
\begin{equation}
    C_e(\bm{r},t) = \frac{1}{L^2} \sum_{\alpha} \left\langle
    {\bm{S}(\bm{r}_\alpha,t) \cdot \bm{S}(\bm{r}_\alpha+\bm{r}},t)
    \right\rangle 
\end{equation} 
with a random initial state at $t=0$.
Figure~\ref{fig6}(a) shows the correlation function, $C_e(\bm{r},t)$,
for $q=8$, $L=4096$, and $K=K_c$.
It decays algebraically with an exponent $\eta$ for small $r$ and 
exponentially for large $r$. The crossover defines the length-scale dependent 
relaxation time $\tau_R$, which turns out to follow a power law 
$\tau_R \sim \xi^{Z_R}$ with the relaxation time exponent $Z_R$.
Figure~\ref{fig6}(b) shows that the correlation functions satisfy 
the scaling form 
\begin{equation}
  C_e(\bm{r},t) = r^{-\eta} F_{e}(r / t^{1/Z_R})  
    \label{C_scaling}
\end{equation}
with $Z_R$ given in Table~\ref{table1}. The correlation function exponent
$\eta$ takes the value around $0.25$ at all values of $q$. 

The relaxation time exponent $Z_R$ characterizes the growth of the 
correlation length $\xi \sim t^{1/Z_R}$ in the transient regime.
It is noteworthy that $Z_R > Z_D$ for all $q$. The particle diffusion is a
faster process than the spin ordering dynamics. 
This result implies that the diffusion-induced spatial heterogeneity
is substantially different from quenched (time-independent) disorder.
Quenched disorder is known to suppress a discontinuous phase
transition~\cite{Greenblatt.2010,Aizenman.1989,Hui.1989}.
For instance, the disordered equilibrium $q$-state Potts model in 2d
has a continuous phase transition at all values of
$q$~\cite{Olson.1999}. The fact that $Z_R >Z_D$ provokes the
question for the mechanism leading to the suppression of the
discontinuous phase transition in the Brownian Potts model with
time-dependent diffusion-induced disorder, which will be addressed shortly.

The correlation time in the steady state can be characterized by the {\em
two-time} correlation function 
\begin{equation}
  C_t(\bm{r},t) = \frac{1}{L^2} \sum_{\alpha} \left\langle
    {\bm{S}(\bm{r}_\alpha,t_0) \cdot \bm{S}(\bm{r}_\alpha+\bm{r}},t_0+t)
    \right\rangle 
\end{equation}
with $t_0 \gg L^{Z_R}$. The spatial correlation decays as the
steady-state fluctuation spreads. This dynamic critical behavior is captured
by the dynamic scaling form
\begin{equation}
  C_t(\bm{r},t) = t^{-\eta/Z_C} F_t ( r / t^{1/Z_C}) 
  \label{Ct_scaling_form}
\end{equation}
with the correlation time exponent $Z_C$ characterizing the correlation 
time scale $\tau_C\sim \xi^{Z_C}$ over a distance $\xi$. 

\begin{figure}[t]
  \includegraphics*[width=\columnwidth]{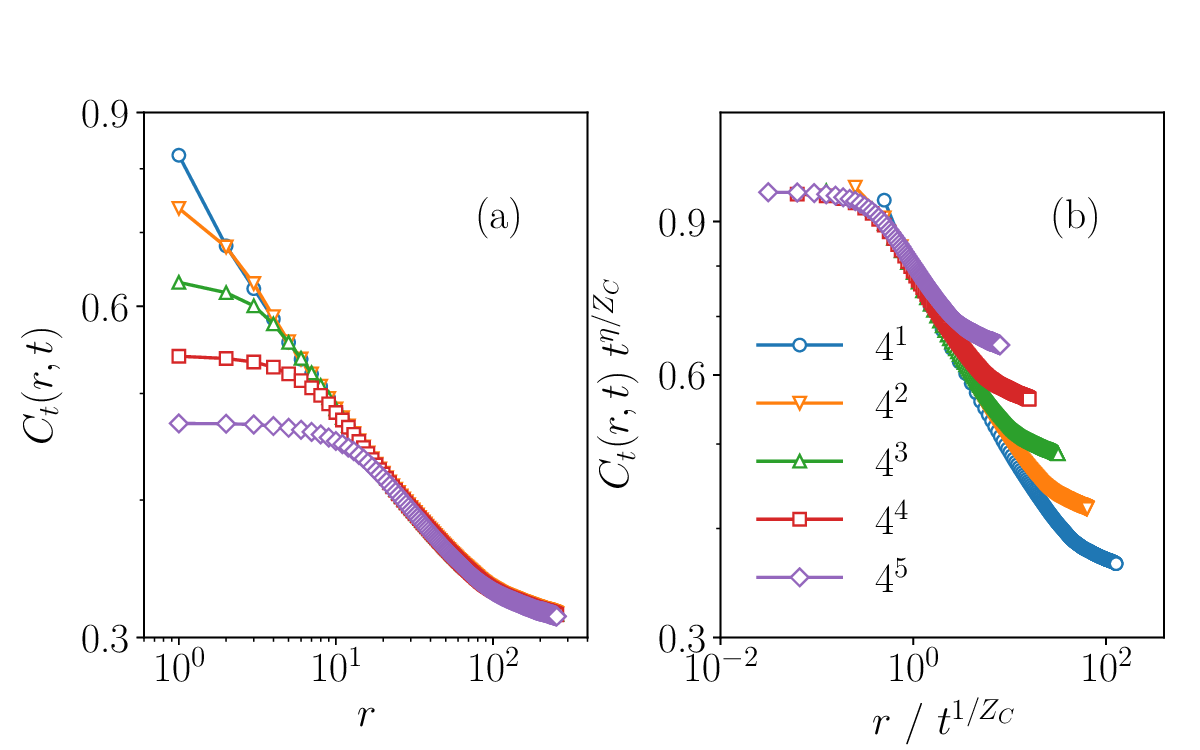}
  \caption{(a) Two-time correlation functions at $t=4^1,\ldots, 4^5$ 
    evaluated at the critical point for $q=3$ and $L=512$ averaged over 100
    samples. 
    (b) Scaling plot according to Eq.~\eqref{Ct_scaling_form} with the
  correlation time exponent $Z_C=2.0$ and $\eta = 0.21$.}
    \label{fig7}
\end{figure}

The numerical data for the critical two-point correlation function $C_t$ for
$q=3$ and $L=512$ are presented in Fig.~\ref{fig7}~\footnote{
The calculation of the two-time correlation function is computationally
demanding since it requires an ensemble average over a large number of 
samples~(100 samples in the present work) 
in the steady state~(i.e. $t_0\gg L^{Z_R}$),
for which reason we restrict ourselves here to $q=3$ and $L\leq 512$. On the
other hand, it is relatively easier to obtain the ensemble-averaged 
equal-time correlation 
function shown in Fig.~\ref{fig6} since it is measured
in the transient regime.}.
Our data for $C_t$ show a good data collapse according to the scaling form \eqref{Ct_scaling_form} with an exponent $Z_C \simeq 2.0$.
This value is close to the dynamic exponent $Z_D$ of diffusion. We expect that $Z_C$
is equal to $Z_D$ for all values of $q$ 
because diffusion is the dominant mechanism that propagates fluctuations.

\begin{figure}
    \includegraphics*[width=0.8\columnwidth]{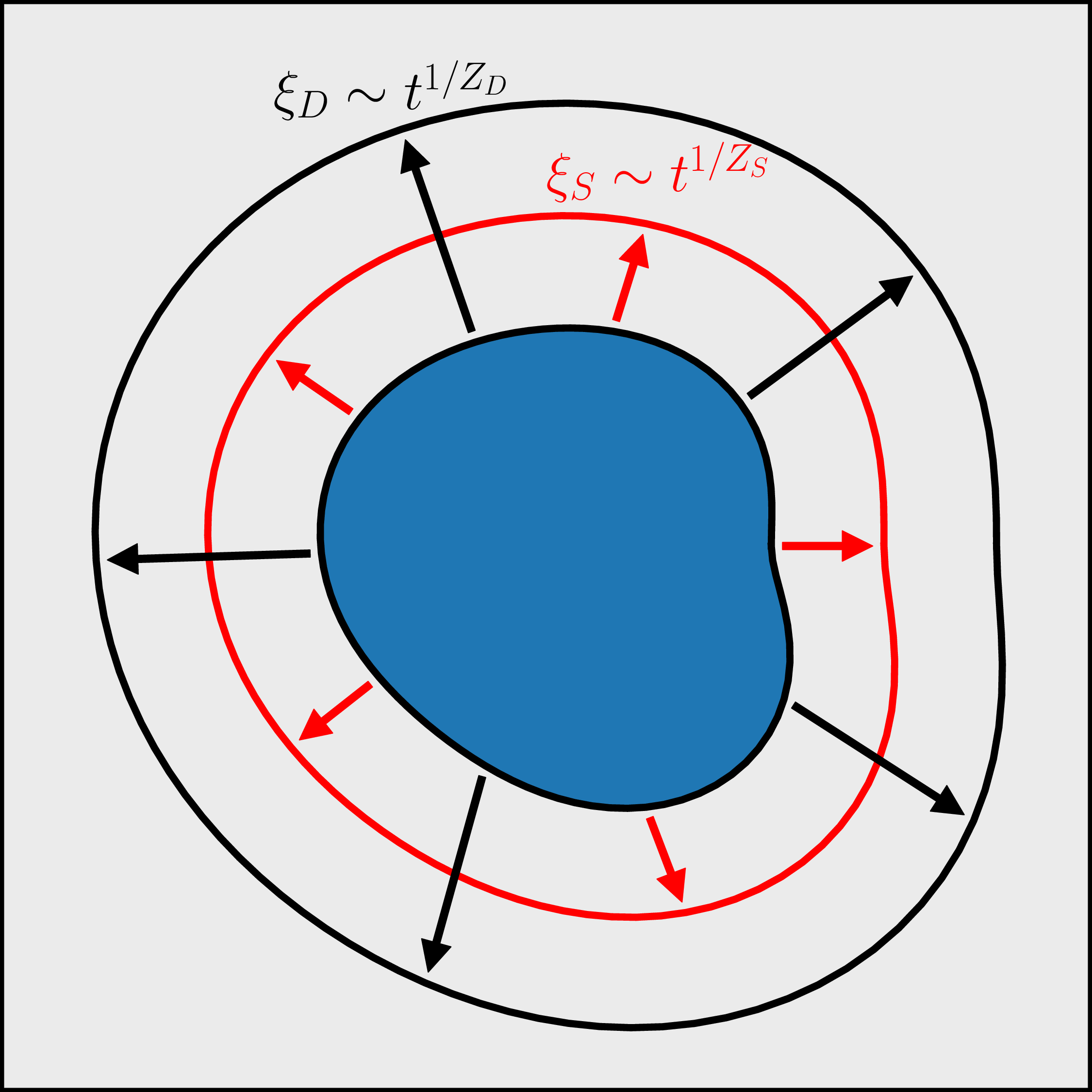}
  \caption{Ferromagnetic domain~(filled blue area) inside a steady state  
  background. The length scales associated with particle diffusion and spin
ordering are represented with the black and red arrows, respectively.} 
  \label{fig8}
\end{figure}

Interestingly, the non-stationary relaxation time 
and the stationary correlation time scale differently
with $Z_R > Z_C \simeq Z_D$. This observation indicates a potential 
reason for the absence of phase coexistence in the Brownian Potts model.
To elaborate we propose a thought-experiment that consists in introducing
a ferromagnetic domain into the system in the steady state at $K=K_c$~(see 
Fig.~\ref{fig8}). The system is then driven away from the steady state
locally. 
Across the domain boundary, ordered spins diffuse into the bulk,
in time $t$ by a distance of $\xi_D \sim t^{1/Z_D}$. On the other hand, 
spin-order propagates in this time only a distance $\xi_R 
\sim t^{1/Z_R} \ll \xi_D$ from the boundary. Thus, the diffusing particles are absorbed into the disordered
bulk and the initially ordered domain keeps shrinking and vanishes eventually.
This argument indicates that particle diffusion destabilizes phase
coexistence as long as $Z_R > Z_D$.
It provides a self-consistent explanation why the Brownian Potts model exhibits a
continuous phase transition without phase coexistence.
We can phrase the same argument in terms of the correlation length $\xi_C \sim
t^{1/Z_C}$: with $Z_C \simeq Z_D < Z_R$, the propagation
of critical fluctuations dominates the domain growth dynamics 
and sweep away ordered domains.

\section{Summary and Discussions}\label{sec:summary}
We have investigated the influence of passive diffusion on the
order-disorder phase transition and the critical behavior in the
Brownian $q$-state Potts model in two dimensions. 
Particle diffusion
introduces interesting aspects that are absent in the equilibrium counterpart 
of immobile spins on a lattice: (i) It breaks detailed balance and drives
the system out of equilibrium. A weak breaking of detailed balance 
is irrelevant for some systems.~\cite{Grinstein.1985,Tauber.2002}. 
In the Brownian Potts model particle diffusion turns out to be strong 
in the sense that it changes the nature of the transition at least for $q>4$. 
(ii) The particle diffusion
introduces time-dependent disorder in the spin-spin interaction network.
At the critical point, we find that density and spin fluctuations 
are faster processes than the ordering dynamics $Z_R > Z_D\simeq Z_C$. 
Thus, the diffusion-induced time-dependent disorder is substantially 
different from quenched disorder.
We argued that time-dependent disorder suppresses
phase coexistence, which is consistent with our numerical results that the
Brownian Potts model displays a continuous phase transition. 

{
    We remark that quenched disorder also suppresses the discontinuity 
    of the phase transition~\cite{Greenblatt.2010}.
    The 2d random-bond Potts models indeed undergo a
    continuous phase transition at any values of $q$~\cite{Cardy.1997,
    Olson.1999}. The correlation length exponent is almost constant
    $\nu\simeq 1.0$ and the correlation function displays a multiscaling
    behavior. It is interesting to note that the correlation length
    exponent of the Brownian Potts model also has a weak dependence on $q$.
    On the other hand, we do not find any evidence for the multiscaling
    behavior in the Brownian Potts model.
}

\begin{figure}
    \includegraphics*[width=\columnwidth]{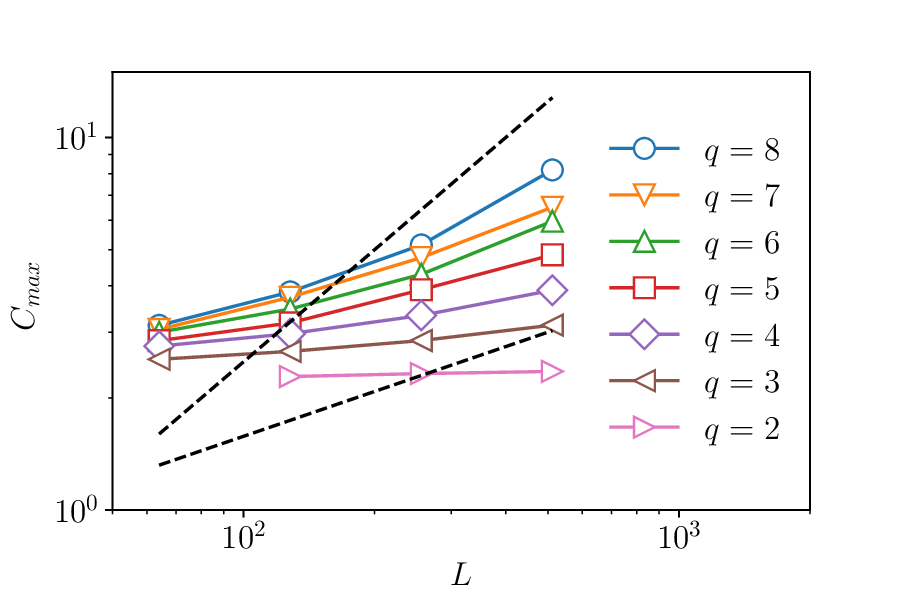}
    \caption{Log-log plots for the peak values of the specific heat. 
      At $q=2$, $C_{max} \sim \ln L$. On the other hand, the curvature for
      $q>3$ indicates a crossover to a power-law scaling. For
      comparison, dashed straight lines with slope $2/5$ 
      and $1$ are shown, which are representative for the equilibrium 
      Potts model with $q=3$ and $4$, respectively.}
    \label{fig9}
\end{figure}

The histograms, shown in Fig.~\ref{fig3}, are the evidence for the absence
of phase coexistence in the Brownian Potts model even for $q>4$. However,
the universality class for the Brownian Potts model remains still elusive.
Although the critical exponents summarized in Table~\ref{table1} vary 
slightly with $q$, the numerical uncertainty is too large to draw a
final conclusion regarding the universality class. 
We also defined an energy-like quantity 
$E=-\frac{1}{2} \sum_{|\bm{r}_i - \bm{r}_j|<r_0} \delta(\sigma_i,\sigma_j)$ 
and measured its second moment $C \equiv (\langle E^2\rangle -
\langle E\rangle^2)/L^2$ as an analogy to the specific heat of the
equilibrium Potts model. Near $K=K_c$, it has a peak whose height increases
as $L$. We present the peak values in Fig.~\ref{fig9}. For $q=2$, it
increases logarithmically with the system size $L$, which is a
characteristics of the Ising universality class. The data for $q\geq 3$
deviate from the logarithmic scaling. The crossover indicates a $q$-dependent
critical behavior.
The curvature in the plot, however, indicates that the asymptotic
behavior can be accessed in much larger systems.

In the Brownian Potts model, particles interact on a time-dependent disordered
network whose edges correspond to particle pairs that have a distance 
$r \le r_0$.
Remarkably, these instantaneous interaction networks do not percolate for the
particle density $\rho=1$ and $r_0=1$ considered here, which
means that they do not contain a single infinite cluster but comprise only
finite clusters of an average size. This can be seen by recurring to 
random plane networks \cite{Gilbert.1961}, which are defined by overlapping
objects of size $a$ distributed randomly in the plane with density $\rho$.
For disks it is known that those networks percolate for $\rho\cdot a\ge 1.127$
\cite{Balister.2005,Mertens.2012},
which implies that for $\rho=1$ the disk radius must be larger than $r_c=0.599$.
For our instantaneous interaction network this means that 
the particle-particle interaction range $r_0$ should be larger than 
$2r_c=1.197$ 
in order to form percolating interaction clusters, which is not the case here.
If the interaction network would be static, i.e. the particle would be immobile,
long range order could not emerge on the basis of only finite interaction
clusters. On the other hand, in the Brownian Potts model long range order 
emerges in spite of the non-percolating instantaneous interaction networks:
Particle diffusion propagates spin order with time 
beyond the instantaneous interaction clusters and thus generates an 
effective, time-averaged global connectivity of the interaction network.
{If the particle density is smaller than a certain threshold
    value $\rho_c$, the interaction network fails to maintain 
    global connectivity and spins cannot order at any coupling constant.
    For the Ising case with $q=2$, we numerically found that the threshold
    density is given by $\rho_c \simeq 0.93$. 
    It would be interesting to study the
    nature of the dynamical percolation transition, which we leave for
future study.}

The coupling between the spin and spatial degrees of freedom 
turns out to be an important aspect determining the nature of phase transitions.
The Brownian Potts model has a unidirectional coupling: Particles diffuse freely
irrespective of spin states, but the particle diffusion modifies the interaction network of spins.
It would be interesting to study the order-disorder transition and the
flocking transition in a system with a coupling in both directions. 
There were a few studies along this line~\cite{OKeeffe.2017,Lee.2021pd8}. We hope 
that our work triggers further systematic analysis of the phase transitions
and the critical phenomena in passively or actively moving spin systems.

\begin{acknowledgments}
  This work is supported by the National Research Foundation of Korea~(KRF)
  grant funded by the Korea government~(MSIP)~[Grant No. 2019R1A2C1009628].
  We acknowledge the computing resources of Urban Big data and AI Institute
  (UBAI) at the University of Seoul.
\end{acknowledgments}

\bibliography{paper}

\begin{thebibliography}{33}%
\makeatletter
\providecommand \@ifxundefined [1]{%
 \@ifx{#1\undefined}
}%
\providecommand \@ifnum [1]{%
 \ifnum #1\expandafter \@firstoftwo
 \else \expandafter \@secondoftwo
 \fi
}%
\providecommand \@ifx [1]{%
 \ifx #1\expandafter \@firstoftwo
 \else \expandafter \@secondoftwo
 \fi
}%
\providecommand \natexlab [1]{#1}%
\providecommand \enquote  [1]{``#1''}%
\providecommand \bibnamefont  [1]{#1}%
\providecommand \bibfnamefont [1]{#1}%
\providecommand \citenamefont [1]{#1}%
\providecommand \href@noop [0]{\@secondoftwo}%
\providecommand \href [0]{\begingroup \@sanitize@url \@href}%
\providecommand \@href[1]{\@@startlink{#1}\@@href}%
\providecommand \@@href[1]{\endgroup#1\@@endlink}%
\providecommand \@sanitize@url [0]{\catcode `\\12\catcode `\$12\catcode
  `\&12\catcode `\#12\catcode `\^12\catcode `\_12\catcode `\%12\relax}%
\providecommand \@@startlink[1]{}%
\providecommand \@@endlink[0]{}%
\providecommand \url  [0]{\begingroup\@sanitize@url \@url }%
\providecommand \@url [1]{\endgroup\@href {#1}{\urlprefix }}%
\providecommand \urlprefix  [0]{URL }%
\providecommand \Eprint [0]{\href }%
\providecommand \doibase [0]{https://doi.org/}%
\providecommand \selectlanguage [0]{\@gobble}%
\providecommand \bibinfo  [0]{\@secondoftwo}%
\providecommand \bibfield  [0]{\@secondoftwo}%
\providecommand \translation [1]{[#1]}%
\providecommand \BibitemOpen [0]{}%
\providecommand \bibitemStop [0]{}%
\providecommand \bibitemNoStop [0]{.\EOS\space}%
\providecommand \EOS [0]{\spacefactor3000\relax}%
\providecommand \BibitemShut  [1]{\csname bibitem#1\endcsname}%
\let\auto@bib@innerbib\@empty
%</preamble>
\bibitem [{\citenamefont {Stanley}(1999)}]{Stanley.1999}%
  \BibitemOpen
  \bibfield  {author} {\bibinfo {author} {\bibfnamefont {H.~E.}\ \bibnamefont
  {Stanley}},\ }\bibfield  {title} {\bibinfo {title} {{Scaling, universality,
  and renormalization: Three pillars of modern critical phenomena}},\
  }\href@noop {} {\bibfield  {journal} {\bibinfo  {journal} {Reviews of Modern
  Physics}\ }\textbf {\bibinfo {volume} {71}},\ \bibinfo {pages} {S358}
  (\bibinfo {year} {1999})}\BibitemShut {NoStop}%
\bibitem [{\citenamefont {Hinrichsen}(2000)}]{Hinrichsen.2000}%
  \BibitemOpen
  \bibfield  {author} {\bibinfo {author} {\bibfnamefont {H.}~\bibnamefont
  {Hinrichsen}},\ }\bibfield  {title} {\bibinfo {title} {{Non-equilibrium
  critical phenomena and phase transitions into absorbing states}},\
  }\href@noop {} {\bibfield  {journal} {\bibinfo  {journal} {Advances in
  Physics}\ }\textbf {\bibinfo {volume} {49}},\ \bibinfo {pages} {815}
  (\bibinfo {year} {2000})}\BibitemShut {NoStop}%
\bibitem [{\citenamefont {Wittkowski}\ \emph {et~al.}(2014)\citenamefont
  {Wittkowski}, \citenamefont {Tiribocchi}, \citenamefont {Stenhammar},
  \citenamefont {Allen}, \citenamefont {Marenduzzo},\ and\ \citenamefont
  {Cates}}]{Wittkowski.2014}%
  \BibitemOpen
  \bibfield  {author} {\bibinfo {author} {\bibfnamefont {R.}~\bibnamefont
  {Wittkowski}}, \bibinfo {author} {\bibfnamefont {A.}~\bibnamefont
  {Tiribocchi}}, \bibinfo {author} {\bibfnamefont {J.}~\bibnamefont
  {Stenhammar}}, \bibinfo {author} {\bibfnamefont {R.~J.}\ \bibnamefont
  {Allen}}, \bibinfo {author} {\bibfnamefont {D.}~\bibnamefont {Marenduzzo}},\
  and\ \bibinfo {author} {\bibfnamefont {M.~E.}\ \bibnamefont {Cates}},\
  }\bibfield  {title} {\bibinfo {title} {{Scalar $\phi^4$ field theory for
  active-particle phase separation}},\ }\href@noop {} {\bibfield  {journal}
  {\bibinfo  {journal} {Nature Communications}\ }\textbf {\bibinfo {volume}
  {5}},\ \bibinfo {pages} {4351} (\bibinfo {year} {2014})}\BibitemShut
  {NoStop}%
\bibitem [{\citenamefont {Ramaswamy}(2017)}]{Ramaswamy.2017}%
  \BibitemOpen
  \bibfield  {author} {\bibinfo {author} {\bibfnamefont {S.}~\bibnamefont
  {Ramaswamy}},\ }\bibfield  {title} {\bibinfo {title} {{Active matter}},\
  }\href@noop {} {\bibfield  {journal} {\bibinfo  {journal} {Journal of
  Statistical Mechanics: Theory and Experiment}\ }\textbf {\bibinfo {volume}
  {2017}},\ \bibinfo {pages} {054002} (\bibinfo {year} {2017})}\BibitemShut
  {NoStop}%
\bibitem [{\citenamefont {Shaebani}\ \emph {et~al.}(2020)\citenamefont
  {Shaebani}, \citenamefont {Wysocki}, \citenamefont {Winkler}, \citenamefont
  {Gompper},\ and\ \citenamefont {Rieger}}]{Shaebani.2020}%
  \BibitemOpen
  \bibfield  {author} {\bibinfo {author} {\bibfnamefont {M.~R.}\ \bibnamefont
  {Shaebani}}, \bibinfo {author} {\bibfnamefont {A.}~\bibnamefont {Wysocki}},
  \bibinfo {author} {\bibfnamefont {R.~G.}\ \bibnamefont {Winkler}}, \bibinfo
  {author} {\bibfnamefont {G.}~\bibnamefont {Gompper}},\ and\ \bibinfo {author}
  {\bibfnamefont {H.}~\bibnamefont {Rieger}},\ }\bibfield  {title} {\bibinfo
  {title} {{Computational models for active matter}},\ }\href@noop {}
  {\bibfield  {journal} {\bibinfo  {journal} {Nature Reviews Physics}\ }\textbf
  {\bibinfo {volume} {2}},\ \bibinfo {pages} {181} (\bibinfo {year}
  {2020})}\BibitemShut {NoStop}%
\bibitem [{\citenamefont {O’Byrne}\ \emph {et~al.}(2022)\citenamefont
  {O’Byrne}, \citenamefont {Kafri}, \citenamefont {Tailleur},\ and\
  \citenamefont {Wijland}}]{OByrne.2022}%
  \BibitemOpen
  \bibfield  {author} {\bibinfo {author} {\bibfnamefont {J.}~\bibnamefont
  {O’Byrne}}, \bibinfo {author} {\bibfnamefont {Y.}~\bibnamefont {Kafri}},
  \bibinfo {author} {\bibfnamefont {J.}~\bibnamefont {Tailleur}},\ and\
  \bibinfo {author} {\bibfnamefont {F.~v.}\ \bibnamefont {Wijland}},\
  }\bibfield  {title} {\bibinfo {title} {{Time irreversibility in active
  matter, from micro to macro}},\ }\href@noop {} {\bibfield  {journal}
  {\bibinfo  {journal} {Nature Reviews Physics}\ }\textbf {\bibinfo {volume}
  {4}},\ \bibinfo {pages} {167} (\bibinfo {year} {2022})}\BibitemShut {NoStop}%
\bibitem [{\citenamefont {Vicsek}\ \emph {et~al.}(1995)\citenamefont {Vicsek},
  \citenamefont {Czirók}, \citenamefont {Jacob}, \citenamefont {Cohen},\ and\
  \citenamefont {Shochet}}]{Vicsek.1995}%
  \BibitemOpen
  \bibfield  {author} {\bibinfo {author} {\bibfnamefont {T.}~\bibnamefont
  {Vicsek}}, \bibinfo {author} {\bibfnamefont {A.}~\bibnamefont {Czirók}},
  \bibinfo {author} {\bibfnamefont {E.~B.}\ \bibnamefont {Jacob}}, \bibinfo
  {author} {\bibfnamefont {I.}~\bibnamefont {Cohen}},\ and\ \bibinfo {author}
  {\bibfnamefont {O.}~\bibnamefont {Shochet}},\ }\bibfield  {title} {\bibinfo
  {title} {{Novel Type of Phase Transition in a System of Self-Driven
  Particles}},\ }\href@noop {} {\bibfield  {journal} {\bibinfo  {journal}
  {Physical Review Letters}\ }\textbf {\bibinfo {volume} {75}},\ \bibinfo
  {pages} {1226 } (\bibinfo {year} {1995})}\BibitemShut {NoStop}%
\bibitem [{\citenamefont {Goldenfeld}(1992)}]{Goldenfeld.1992}%
  \BibitemOpen
  \bibfield  {author} {\bibinfo {author} {\bibfnamefont {N.}~\bibnamefont
  {Goldenfeld}},\ }\href@noop {} {\emph {\bibinfo {title} {{Lectures on phase
  transitions and the renormalization group}}}},\ Addison-Wesley\ (\bibinfo
  {publisher} {Addison-Wesley},\ \bibinfo {year} {1992})\BibitemShut {NoStop}%
\bibitem [{\citenamefont {Cates}\ and\ \citenamefont {Tailleur}(5
  03)}]{Cates.2014}%
  \BibitemOpen
  \bibfield  {author} {\bibinfo {author} {\bibfnamefont {M.~E.}\ \bibnamefont
  {Cates}}\ and\ \bibinfo {author} {\bibfnamefont {J.}~\bibnamefont
  {Tailleur}},\ }\bibfield  {title} {\bibinfo {title} {{Motility-Induced Phase
  Separation}},\ }\href@noop {} {\bibfield  {journal} {\bibinfo  {journal}
  {Annual Review of Condensed Matter Physics}\ }\textbf {\bibinfo {volume}
  {6}},\ \bibinfo {pages} {1} (\bibinfo {year} {2015-03})}\BibitemShut
  {NoStop}%
\bibitem [{\citenamefont {Solon}\ and\ \citenamefont {Tailleur}(3
  08)}]{Solon.2013}%
  \BibitemOpen
  \bibfield  {author} {\bibinfo {author} {\bibfnamefont {A.~P.}\ \bibnamefont
  {Solon}}\ and\ \bibinfo {author} {\bibfnamefont {J.}~\bibnamefont
  {Tailleur}},\ }\bibfield  {title} {\bibinfo {title} {{Revisiting the Flocking
  Transition Using Active Spins}},\ }\href@noop {} {\bibfield  {journal}
  {\bibinfo  {journal} {Physical Review Letters}\ }\textbf {\bibinfo {volume}
  {111}},\ \bibinfo {pages} {078101} (\bibinfo {year} {2013-08})}\BibitemShut
  {NoStop}%
\bibitem [{\citenamefont {Solon}\ and\ \citenamefont
  {Tailleur}(2015)}]{Solon.2015f1h}%
  \BibitemOpen
  \bibfield  {author} {\bibinfo {author} {\bibfnamefont {A.~P.}\ \bibnamefont
  {Solon}}\ and\ \bibinfo {author} {\bibfnamefont {J.}~\bibnamefont
  {Tailleur}},\ }\bibfield  {title} {\bibinfo {title} {{Flocking with discrete
  symmetry: The two-dimensional active Ising model}},\ }\href@noop {}
  {\bibfield  {journal} {\bibinfo  {journal} {Physical Review E}\ }\textbf
  {\bibinfo {volume} {92}},\ \bibinfo {pages} {042119} (\bibinfo {year}
  {2015})}\BibitemShut {NoStop}%
\bibitem [{\citenamefont {Chatterjee}\ \emph {et~al.}(2020)\citenamefont
  {Chatterjee}, \citenamefont {Mangeat}, \citenamefont {Paul},\ and\
  \citenamefont {Rieger}}]{Chatterjee.2020}%
  \BibitemOpen
  \bibfield  {author} {\bibinfo {author} {\bibfnamefont {S.}~\bibnamefont
  {Chatterjee}}, \bibinfo {author} {\bibfnamefont {M.}~\bibnamefont {Mangeat}},
  \bibinfo {author} {\bibfnamefont {R.}~\bibnamefont {Paul}},\ and\ \bibinfo
  {author} {\bibfnamefont {H.}~\bibnamefont {Rieger}},\ }\bibfield  {title}
  {\bibinfo {title} {{Flocking and reorientation transition in the 4-state
  active Potts model}},\ }\href@noop {} {\bibfield  {journal} {\bibinfo
  {journal} {EPL}\ }\textbf {\bibinfo {volume} {130}},\ \bibinfo {pages}
  {66001} (\bibinfo {year} {2020})}\BibitemShut {NoStop}%
\bibitem [{\citenamefont {Mangeat}\ \emph {et~al.}(2020)\citenamefont
  {Mangeat}, \citenamefont {Chatterjee}, \citenamefont {Paul},\ and\
  \citenamefont {Rieger}}]{Mangeat.2020}%
  \BibitemOpen
  \bibfield  {author} {\bibinfo {author} {\bibfnamefont {M.}~\bibnamefont
  {Mangeat}}, \bibinfo {author} {\bibfnamefont {S.}~\bibnamefont {Chatterjee}},
  \bibinfo {author} {\bibfnamefont {R.}~\bibnamefont {Paul}},\ and\ \bibinfo
  {author} {\bibfnamefont {H.}~\bibnamefont {Rieger}},\ }\bibfield  {title}
  {\bibinfo {title} {{Flocking with a q-fold discrete symmetry: Band-to-lane
  transition in the active Potts model}},\ }\href@noop {} {\bibfield  {journal}
  {\bibinfo  {journal} {Physical Review E}\ }\textbf {\bibinfo {volume}
  {102}},\ \bibinfo {pages} {042601} (\bibinfo {year} {2020})}\BibitemShut
  {NoStop}%
\bibitem [{\citenamefont {Wu}(1982)}]{Wu.1982k4e}%
  \BibitemOpen
  \bibfield  {author} {\bibinfo {author} {\bibfnamefont {F.~Y.}\ \bibnamefont
  {Wu}},\ }\bibfield  {title} {\bibinfo {title} {{The Potts model}},\
  }\href@noop {} {\bibfield  {journal} {\bibinfo  {journal} {Reviews of Modern
  Physics}\ }\textbf {\bibinfo {volume} {54}},\ \bibinfo {pages} {235}
  (\bibinfo {year} {1982})}\BibitemShut {NoStop}%
\bibitem [{Note1()}]{Note1}%
  \BibitemOpen
  \bibinfo {note} {Since we adopt the parallel update rule, the spins relax
  into a thermal equilibrium state associated with a Hamiltonian that is
  slightly modified from the conventional Potts model Hamiltonian.}\BibitemShut
  {Stop}%
\bibitem [{\citenamefont {Park}\ and\ \citenamefont {Noh}(2017)}]{Park.2017}%
  \BibitemOpen
  \bibfield  {author} {\bibinfo {author} {\bibfnamefont {J.-M.}\ \bibnamefont
  {Park}}\ and\ \bibinfo {author} {\bibfnamefont {J.~D.}\ \bibnamefont {Noh}},\
  }\bibfield  {title} {\bibinfo {title} {{Tricritical behavior of
  nonequilibrium Ising spins in fluctuating environments}},\ }\href@noop {}
  {\bibfield  {journal} {\bibinfo  {journal} {Physical Review E}\ }\textbf
  {\bibinfo {volume} {95}},\ \bibinfo {pages} {042106} (\bibinfo {year}
  {2017})}\BibitemShut {NoStop}%
\bibitem [{\citenamefont {Jȩdrzejewski}\ \emph {et~al.}(2015)\citenamefont
  {Jȩdrzejewski}, \citenamefont {Chmiel},\ and\ \citenamefont
  {Sznajd-Weron}}]{Jedrzejewski.2015}%
  \BibitemOpen
  \bibfield  {author} {\bibinfo {author} {\bibfnamefont {A.}~\bibnamefont
  {Jȩdrzejewski}}, \bibinfo {author} {\bibfnamefont {A.}~\bibnamefont
  {Chmiel}},\ and\ \bibinfo {author} {\bibfnamefont {K.}~\bibnamefont
  {Sznajd-Weron}},\ }\bibfield  {title} {\bibinfo {title} {{Oscillating
  hysteresis in the q-neighbor Ising model}},\ }\href@noop {} {\bibfield
  {journal} {\bibinfo  {journal} {Physical Review E}\ }\textbf {\bibinfo
  {volume} {92}},\ \bibinfo {pages} {052105} (\bibinfo {year}
  {2015})}\BibitemShut {NoStop}%
\bibitem [{\citenamefont {Imry}\ and\ \citenamefont {Ma}(1975)}]{Imry.1975}%
  \BibitemOpen
  \bibfield  {author} {\bibinfo {author} {\bibfnamefont {Y.}~\bibnamefont
  {Imry}}\ and\ \bibinfo {author} {\bibfnamefont {S.-k.}\ \bibnamefont {Ma}},\
  }\bibfield  {title} {\bibinfo {title} {{Random-Field Instability of the
  Ordered State of Continuous Symmetry}},\ }\href@noop {} {\bibfield  {journal}
  {\bibinfo  {journal} {Physical Review Letters}\ }\textbf {\bibinfo {volume}
  {35}},\ \bibinfo {pages} {1399} (\bibinfo {year} {1975})}\BibitemShut
  {NoStop}%
\bibitem [{\citenamefont {Greenblatt}\ \emph {et~al.}(2010)\citenamefont
  {Greenblatt}, \citenamefont {Aizenman},\ and\ \citenamefont
  {Lebowitz}}]{Greenblatt.2010}%
  \BibitemOpen
  \bibfield  {author} {\bibinfo {author} {\bibfnamefont {R.~L.}\ \bibnamefont
  {Greenblatt}}, \bibinfo {author} {\bibfnamefont {M.}~\bibnamefont
  {Aizenman}},\ and\ \bibinfo {author} {\bibfnamefont {J.~L.}\ \bibnamefont
  {Lebowitz}},\ }\bibfield  {title} {\bibinfo {title} {{On spin systems with
  quenched randomness: Classical and quantum}},\ }\href@noop {} {\bibfield
  {journal} {\bibinfo  {journal} {Physica A: Statistical Mechanics and its
  Applications}\ }\textbf {\bibinfo {volume} {389}},\ \bibinfo {pages} {2902}
  (\bibinfo {year} {2010})}\BibitemShut {NoStop}%
\bibitem [{\citenamefont {Aizenman}\ and\ \citenamefont
  {Wehr}(1989)}]{Aizenman.1989}%
  \BibitemOpen
  \bibfield  {author} {\bibinfo {author} {\bibfnamefont {M.}~\bibnamefont
  {Aizenman}}\ and\ \bibinfo {author} {\bibfnamefont {J.}~\bibnamefont
  {Wehr}},\ }\bibfield  {title} {\bibinfo {title} {{Rounding of First-Order
  Phase Transitions in Systems with Quenched Disorder}},\ }\href@noop {}
  {\bibfield  {journal} {\bibinfo  {journal} {Physical Review Letters}\
  }\textbf {\bibinfo {volume} {62}},\ \bibinfo {pages} {2503} (\bibinfo {year}
  {1989})}\BibitemShut {NoStop}%
\bibitem [{\citenamefont {Hui}\ and\ \citenamefont {Berker}(1989)}]{Hui.1989}%
  \BibitemOpen
  \bibfield  {author} {\bibinfo {author} {\bibfnamefont {K.}~\bibnamefont
  {Hui}}\ and\ \bibinfo {author} {\bibfnamefont {A.~N.}\ \bibnamefont
  {Berker}},\ }\bibfield  {title} {\bibinfo {title} {{Random-field mechanism in
  random-bond multicritical systems}},\ }\href@noop {} {\bibfield  {journal}
  {\bibinfo  {journal} {Physical Review Letters}\ }\textbf {\bibinfo {volume}
  {62}},\ \bibinfo {pages} {2507} (\bibinfo {year} {1989})}\BibitemShut
  {NoStop}%
\bibitem [{\citenamefont {Olson}\ and\ \citenamefont
  {Young}(1999)}]{Olson.1999}%
  \BibitemOpen
  \bibfield  {author} {\bibinfo {author} {\bibfnamefont {T.}~\bibnamefont
  {Olson}}\ and\ \bibinfo {author} {\bibfnamefont {A.~P.}\ \bibnamefont
  {Young}},\ }\bibfield  {title} {\bibinfo {title} {{Monte Carlo study of the
  critical behavior of random bond Potts models}},\ }\href@noop {} {\bibfield
  {journal} {\bibinfo  {journal} {Physical Review B}\ }\textbf {\bibinfo
  {volume} {60}},\ \bibinfo {pages} {3428} (\bibinfo {year}
  {1999})}\BibitemShut {NoStop}%
\bibitem [{\citenamefont {Newman}\ and\ \citenamefont
  {Barkema}(1999)}]{newman1999monte}%
  \BibitemOpen
  \bibfield  {author} {\bibinfo {author} {\bibfnamefont {M.}~\bibnamefont
  {Newman}}\ and\ \bibinfo {author} {\bibfnamefont {G.~T.}\ \bibnamefont
  {Barkema}},\ }\href@noop {} {\emph {\bibinfo {title} {{Monte Carlo Methods in
  Statistical Physics}}}}\ (\bibinfo  {publisher} {Oxford University Press},\
  \bibinfo {address} {New York},\ \bibinfo {year} {1999})\BibitemShut {NoStop}%
\bibitem [{\citenamefont {Binder}\ and\ \citenamefont
  {Heermann}(2010)}]{Binder.1997}%
  \BibitemOpen
  \bibfield  {author} {\bibinfo {author} {\bibfnamefont {K.}~\bibnamefont
  {Binder}}\ and\ \bibinfo {author} {\bibfnamefont {D.~W.}\ \bibnamefont
  {Heermann}},\ }\href@noop {} {\emph {\bibinfo {title} {{Monte Carlo
  Simulation in Statistical Physics, An Introduction}}}},\ \bibinfo {series}
  {Springer Series in Solid-State Sciences}, Vol.~\bibinfo {volume} {80}\
  (\bibinfo  {publisher} {Springer-Verlag},\ \bibinfo {year}
  {2010})\BibitemShut {NoStop}%
\bibitem [{Note2()}]{Note2}%
  \BibitemOpen
  \bibinfo {note} {The calculation of the two-time correlation function is
  computationally demanding since it requires an ensemble average over a large
  number of samples~(100 samples in the present work) in the steady state~(i.e.
  $t_0\gg L^{Z_R}$), for which reason we restrict ourselves here to $q=3$ and
  $L\leq 512$. On the other hand, it is relatively easier to obtain the
  ensemble-averaged equal-time correlation function shown in Fig.~\ref {fig6}
  since it is measured in the transient regime.}\BibitemShut {Stop}%
\bibitem [{\citenamefont {Grinstein}\ \emph {et~al.}(1985)\citenamefont
  {Grinstein}, \citenamefont {Jayaprakash},\ and\ \citenamefont
  {He}}]{Grinstein.1985}%
  \BibitemOpen
  \bibfield  {author} {\bibinfo {author} {\bibfnamefont {G.}~\bibnamefont
  {Grinstein}}, \bibinfo {author} {\bibfnamefont {C.}~\bibnamefont
  {Jayaprakash}},\ and\ \bibinfo {author} {\bibfnamefont {Y.}~\bibnamefont
  {He}},\ }\bibfield  {title} {\bibinfo {title} {{Statistical Mechanics of
  Probabilistic Cellular Automata}},\ }\href@noop {} {\bibfield  {journal}
  {\bibinfo  {journal} {Physical Review Letters}\ }\textbf {\bibinfo {volume}
  {55}},\ \bibinfo {pages} {2527} (\bibinfo {year} {1985})}\BibitemShut
  {NoStop}%
\bibitem [{\citenamefont {Täuber}\ \emph {et~al.}(2002)\citenamefont
  {Täuber}, \citenamefont {Akkineni},\ and\ \citenamefont
  {Santos}}]{Tauber.2002}%
  \BibitemOpen
  \bibfield  {author} {\bibinfo {author} {\bibfnamefont {U.~C.}\ \bibnamefont
  {Täuber}}, \bibinfo {author} {\bibfnamefont {V.~K.}\ \bibnamefont
  {Akkineni}},\ and\ \bibinfo {author} {\bibfnamefont {J.~E.}\ \bibnamefont
  {Santos}},\ }\bibfield  {title} {\bibinfo {title} {{Effects of Violating
  Detailed Balance on Critical Dynamics}},\ }\href@noop {} {\bibfield
  {journal} {\bibinfo  {journal} {Physical Review Letters}\ }\textbf {\bibinfo
  {volume} {88}},\ \bibinfo {pages} {045702} (\bibinfo {year}
  {2002})}\BibitemShut {NoStop}%
\bibitem [{\citenamefont {Cardy}\ and\ \citenamefont
  {Jacobsen}(1997)}]{Cardy.1997}%
  \BibitemOpen
  \bibfield  {author} {\bibinfo {author} {\bibfnamefont {J.}~\bibnamefont
  {Cardy}}\ and\ \bibinfo {author} {\bibfnamefont {J.~L.}\ \bibnamefont
  {Jacobsen}},\ }\bibfield  {title} {\bibinfo {title} {{Critical Behavior of
  Random-Bond Potts Models}},\ }\href
  {https://doi.org/10.1103/physrevlett.79.4063} {\bibfield  {journal} {\bibinfo
   {journal} {Physical Review Letters}\ }\textbf {\bibinfo {volume} {79}},\
  \bibinfo {pages} {4063} (\bibinfo {year} {1997})}\BibitemShut {NoStop}%
\bibitem [{\citenamefont {Gilbert}(1961)}]{Gilbert.1961}%
  \BibitemOpen
  \bibfield  {author} {\bibinfo {author} {\bibfnamefont {E.~N.}\ \bibnamefont
  {Gilbert}},\ }\bibfield  {title} {\bibinfo {title} {{Random Plane
  Networks}},\ }\href@noop {} {\bibfield  {journal} {\bibinfo  {journal}
  {Journal of the Society for Industrial and Applied Mathematics}\ }\textbf
  {\bibinfo {volume} {9}},\ \bibinfo {pages} {533} (\bibinfo {year}
  {1961})}\BibitemShut {NoStop}%
\bibitem [{\citenamefont {Balister}\ \emph {et~al.}(2005)\citenamefont
  {Balister}, \citenamefont {Bollobás},\ and\ \citenamefont
  {Walters}}]{Balister.2005}%
  \BibitemOpen
  \bibfield  {author} {\bibinfo {author} {\bibfnamefont {P.}~\bibnamefont
  {Balister}}, \bibinfo {author} {\bibfnamefont {B.}~\bibnamefont
  {Bollobás}},\ and\ \bibinfo {author} {\bibfnamefont {M.}~\bibnamefont
  {Walters}},\ }\bibfield  {title} {\bibinfo {title} {{Continuum percolation
  with steps in the square or the disc}},\ }\href@noop {} {\bibfield  {journal}
  {\bibinfo  {journal} {Random Structures \& Algorithms}\ }\textbf {\bibinfo
  {volume} {26}},\ \bibinfo {pages} {392} (\bibinfo {year} {2005})}\BibitemShut
  {NoStop}%
\bibitem [{\citenamefont {Mertens}\ and\ \citenamefont
  {Moore}(2012)}]{Mertens.2012}%
  \BibitemOpen
  \bibfield  {author} {\bibinfo {author} {\bibfnamefont {S.}~\bibnamefont
  {Mertens}}\ and\ \bibinfo {author} {\bibfnamefont {C.}~\bibnamefont
  {Moore}},\ }\bibfield  {title} {\bibinfo {title} {{Continuum percolation
  thresholds in two dimensions}},\ }\href@noop {} {\bibfield  {journal}
  {\bibinfo  {journal} {Physical Review E}\ }\textbf {\bibinfo {volume} {86}},\
  \bibinfo {pages} {061109} (\bibinfo {year} {2012})}\BibitemShut {NoStop}%
\bibitem [{\citenamefont {O’Keeffe}\ \emph {et~al.}(2017)\citenamefont
  {O’Keeffe}, \citenamefont {Hong},\ and\ \citenamefont
  {Strogatz}}]{OKeeffe.2017}%
  \BibitemOpen
  \bibfield  {author} {\bibinfo {author} {\bibfnamefont {K.~P.}\ \bibnamefont
  {O’Keeffe}}, \bibinfo {author} {\bibfnamefont {H.}~\bibnamefont {Hong}},\
  and\ \bibinfo {author} {\bibfnamefont {S.~H.}\ \bibnamefont {Strogatz}},\
  }\bibfield  {title} {\bibinfo {title} {{Oscillators that sync and swarm}},\
  }\href@noop {} {\bibfield  {journal} {\bibinfo  {journal} {Nature
  Communications}\ }\textbf {\bibinfo {volume} {8}},\ \bibinfo {pages} {1504}
  (\bibinfo {year} {2017})}\BibitemShut {NoStop}%
\bibitem [{\citenamefont {Lee}\ \emph {et~al.}(2021)\citenamefont {Lee},
  \citenamefont {Yeo},\ and\ \citenamefont {Hong}}]{Lee.2021pd8}%
  \BibitemOpen
  \bibfield  {author} {\bibinfo {author} {\bibfnamefont {H.~K.}\ \bibnamefont
  {Lee}}, \bibinfo {author} {\bibfnamefont {K.}~\bibnamefont {Yeo}},\ and\
  \bibinfo {author} {\bibfnamefont {H.}~\bibnamefont {Hong}},\ }\bibfield
  {title} {\bibinfo {title} {{Collective steady-state patterns of swarmalators
  with finite-cutoff interaction distance}},\ }\href@noop {} {\bibfield
  {journal} {\bibinfo  {journal} {Chaos: An Interdisciplinary Journal of
  Nonlinear Science}\ }\textbf {\bibinfo {volume} {31}},\ \bibinfo {pages}
  {033134} (\bibinfo {year} {2021})}\BibitemShut {NoStop}%
\end{thebibliography}%
\end{document}